# On the Relation between Centrality Measures and Consensus Algorithms

Amir Noori

*Sama Technical and Vocational Training College, Islamic Azad University, Karaj branch, Karaj, Iran*
*E-mail*: amir.noori@kiau.ac.ir

**ABSTRACT**

*This paper introduces some tools from graph theory and distributed consensus algorithms to construct an optimal, yet robust, hierarchical information sharing structure for large-scale decision making and control problems. The proposed method is motivated by the robustness and optimality of leaf-venation patterns. We introduce a new class of centrality measures which are built based on the degree distribution of nodes within network graph. Furthermore, the proposed measure is used to select the appropriate weight of the corresponding consensus algorithm. To this end, an implicit hierarchical structure is derived that control the flow of information in different situations. In addition, the performance analysis of the proposed measure with respect to other standard measures is performed to investigate the convergence and asymptotic behavior of the measure. Gas Transmission Network is served as our test-bed to demonstrate the applicability and the efficiently of the method.*

## 1. INTRODUCTION

In large scale systems, resources are distributed over a wide geographical area and it is important not only to maintain network safe performance in the face of disasters, but also to make some decisions to optimize its performance. In such circumstance, each local decision maker (which is represented by *nodes* in graph theory context and by *agents* in a multi-agent setting) requires both local and appropriate global information to make optimal decisions. Naturally, units with large mass transportation capability and/or with extra information links have leadership ability and create several clusters. In particular, communications between leaders of clusters may create a scale-free network topology.

From social network perspective, it is important to detect the role and the relative importance of individual decision makers within networks. It helps decision makers to process, re-transmit and direct appropriate information to optimal and robust paths in the network. In fact, performance of distributed algorithms largely relies on structures that are used to share the information within the network. In distributed systems, finding optimal yet robust information sharing structures (and related protocols) is a challenging problem.

In graph theory, the relative importance of a *vertex* or *edge* within a graph is determined using *Centrality Measures*. The node degree and the shortest path between nodes are the most informative values available to determine advantage of a node with respect to its neighbors. Degree, betweeness, closeness (as local measures) and eigenvector (as global or *spectral* measure) are the most commonly used centrality measures [3]. Eigenvector and similar centrality measures have advantages over graph-theoretic measures like degree, betweeness, and closeness; they can be used in signed and valued graphs [4]. Extension of local measure to also include these types of graphs is in progress. However, the calculation of these measures often requires global knowledge of the network properties and network topology. This problem limits the applicability of these measures to distributed systems which necessitate scalability and make extensive use of local information.

measures to distributed systems which necessitate scalability and make extensive use of local information.

Some centrality measures make implicit assumptions about the manner in which traffic flows through a network and most commonly used centrality measures are not appropriate for most of the flows we are routinely interested in [1]. In [2], it is explained how betweeness centrality measure implicitly assumes that information spreads only along shortest paths and then propose a betweeness measure that relaxes this assumption while include contributions from essentially all paths between nodes, not just the shortest, although it still gives more weight to short paths.

In this paper, we introduce a novel eigenvector-like centrality measure, called *Cumulative Degree (CD)* and *Distributed Cumulative Degree (DCD)* measure. The idea is simple; this measure works on degree distribution of nodes within network where considers the effect of all other reachable nodes in the network graph that are accessible through spanning tree with (or without) a weighting or directing mechanism. It can be shown that the recursive computation of this measure is possible. In large problems and for computational tractability, we show that this measure can be approximated by truncated spanning trees with desired accuracy. Further, we show that the proposed measure have several desired properties like standard measures; eigenvector and betweeness measure and so on. In particular, through several simulations and applications, we analyze its performance. To this end, we show that the proposed measure can provide good estimate of the corresponding optimal flow. We also investigate different network flow with respect to network *Optimality* and *Robustness* issues. In every transportation network, nodes in shortest paths should be strengthen to achieve optimal flow, while for robustness and fault tolerance issues other links should be taken into account. Using this measure every decision maker may exploit its estimate of power of its neighbors or routing paths. In addition, we introduce *Directed Consensus* (DC) algorithm which is a kind of consensus seeking algorithms in which its weight selection mechanism is performed based on Cumulative Degree measure. Through simulations we show that the proposed measure has more flexibility and scalability than eigenvector centrality measure. In particular, it works based on the well-known scoring principle that connections to high-scoring nodes contribute more to the score of the node in question than equal connections to low-scoring nodes that resemble importance and popularity.

The remainder of the work is organized as follows; the following section presents some preliminaries as well as notational conventions from graph theory and centrality measures. In the section three, distributed centrality measures, performance analysis and related consensus algorithms are discussed. Section four is devoted to presents application of the measure to the optimization and crisis management in *Gas Transmission Network*. Section six concludes the paper.

## 2. PRELIMINARIES

The framework presented in this paper relies on some basic concepts in graph theory and centrality measures and consensus algorithms that are discussed in the following.

### 2.1. Graph Theory and Centrality Measures

In distributed systems, many problems of practical interests can be represented by graphs. A graph is an ordered pair $G = (V, E)$, consists of a set V of vertices or nodes and a set $E \subseteq (V \times V)$ of edges or lines. Neighbors of node m is defined as

$$N_m = \{i, (m, i) \in E\}$$

From a graph theoretic perspective, the relative importance of a vertex or edge within a graph is determined using centrality measures. At a more substantive level, measures of centrality summarize a node's involvement in or contribution to the cohesiveness of the network [5]. Centrality measures have many applications in social networks [6]. Cook et al [7] have shown that power is not same as centrality in exchange networks. However, this example necessitates careful considerations according to related domain of interest as well as type of network flow (as cited in previous section). A formal definition of the most widely used centrality measures in network analysis are adopted from [5][6][8], which include degree, betweeness, closeness and eigenvector centrality measure.

**Definition** Degree centrality, is the count of the degree or number of adjacencies for a vertex, *v*(m) over its maximum value, *i.e.*, N-1:

$$C_D(m) = \frac{\sum_{i=1}^{N} a(v(i), v(m))}{N-1} = \frac{\deg(m)}{N-1}$$

Where N is the number of vertices or network size and $a(v(i), v(m))$ is the corresponding element of adjacency matrix of the network graph $G = (V, E)$.

**Definition** Betweeness centrality is an index for the number of times in which a vertex occurs on shortest paths between other vertices and is defined by following formula

$$C_B(m) = \sum_{s \neq m \neq t \in V} \frac{\sigma_{st}(m)}{\sigma_{st}}$$

Where $\sigma_{st}$ is the number of shortest paths from $s$ to $t$, and $\sigma_{st}(m)$ is the number of shortest paths from $s$ to $t$ that pass through a vertex $m$. However, computation of this centrality for all vertices is a time-consuming process and involves calculation of the shortest paths between all pairs of vertices on a graph.

**Definition** Closeness centrality is defined as the overall mean of the shortest path between a vertex $m$ and all other vertices reachable from it

$$C_c(m) = \frac{\sum_{t \in C \setminus m} d_G(m,t)}{M-1}$$

Where M is the size of the network's connectivity component C reachable from $m$.

**Definition** Let $x_i$ denote the score of the $i^{th}$ node and $A_{ij}$ be the adjacency matrix of the network. Then, the Eigenvector centrality for a node $i$ is defined as follow

$$x_i = \frac{1}{\lambda} \sum_{j \in N_i} x_j = \frac{1}{\lambda} \sum_{j=1}^{N} A_{ij} x_j$$

In general, the A's entries can be real numbers that represent the related connection strength, as in a signed or valued graph. Eigenvector centrality is a measure of the importance of a node in a network. It assigns relative scores to all nodes in the network based on the well-known principle that connections to high-scoring nodes contribute more to the score of the node in question than equal connections to low-scoring nodes.

In the next section, we will show that the proposed centrality measure inherits several desired properties of both node and spectral centrality measures and it is adjustable according to the selected discounting series and its parameters.

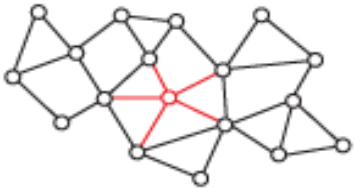

**Fig. 1 graphical representation of closeness and betweeness centrality measures**

## 2.2. Distributed Consensus Algorithms

Information sharing plays a crucial role in every distributed system. In recent years, with the aim of growing public communication systems, it is possible to connect every subsystem of a large scale distributed system. This framework provides prerequisites for acquiring information and applying required control actions. Yet, the key questions about effective communication structure and information distribution not answered completely, especially when switching is allowed in communication structure and agents' dynamic changes.

$$N_i = \{j; (i,j) \in E\}$$

(Or may be some other nodes) and also find some simple aggregation rules like

$$x_i(t+1) = w_{ii} x_i(t) + \sum_{j \in N_i} w_{ij} x_j(t)$$

to reach a consensus on $x$ between all agents

$$\lim_{t \to \infty} [x_i(t) - x_j(t)] \to 0 \quad i,j = 1,2,...,n$$

or

$$\lim_{t \to \infty} x_i(t) \to \frac{1}{N} \sum_{j=1}^{n} x_j(0) \quad i,j = 1,2,...,n \quad (1\text{-}1)$$

Equation (1-1) puts emphasis on average aggregation by parties over initial conditions. From different theoretical development issues of this problem, design of efficient weight factors has received more attention. The effect of weights are crucial on the aggregated value, convergence rate and robustness especially in the situations which changes on topology occurs (e.g., link failure, controlled switching or etc.). In [9], the following rule for weight selection is proposed

$$w_{ij}(t) = \begin{cases} 1/(1+d_i(t)) & (i,j) \in E \text{ or } i=j \\ 0 & \text{otherwise} \end{cases}$$

Where $d_i(t) = |N_i|$ is the degree of node i. However, this choice of weights does not preserve averaging (*i.e.*, eq. 1-1). In [10], a well-known weight design algorithm is proposed that preserves averaging

$$w_{ij}(t) = \begin{cases} 1/(1+\max(d_i(t),d_j(t))) & (i,j) \in E \\ 1 - \sum_{l \in N_i} w_{il} & i=j \\ 0 & \text{otherwise} \end{cases}$$

Maximum-degree weights is another weighted consensus algorithm that preserves averaging and is proposed in [11],

$$w_{ij}(t) = \begin{cases} 1/N & (i,j) \in E \\ 1 - d_i/N & i = j \\ 0 & \text{otherwise} \end{cases}$$

In [12], it is shown that the graphs with small-world property have more faster mixing time than similar regular or random graphs, so are more optimal. In the next section, we propose a bio-inspired information sharing architecture that provides a trade-off between robustness and optimality issues for related consensus algorithm.

## 3. DISTRIBUTED CENTRALITY MEASURES

In network analysis, the power of individual units is not an individual attribute, but arises from their interactions with others. In a more abstract level and only based on the connection between units, these network-wide relations may represent different forms of powers; higher connections or closer to other units, more central unit within network. In section II, we have summarized some corresponding measures which in graph theoretic context respectively called; degree, closeness and betweeness centrality measures. There are several graph theoretic attributes like degree and path to score a specific network flow or degree importance.

Another important class of centrality measures can be defined based on the corresponding adjacent matrix of the underlying network using *Algebraic Graph Theory* [13], which usually called spectral measures like eigenvector centrality measure. These measures reveal some unique properties of nodes in the cost of requirement of network global knowledge. This problem limits the applicability of these measures in distributed algorithms. Here, we start from the well-known node degree measure and generalize it to a broader sense.

### 3.1. Distributed Centrality Measures: Definitions

In this section, we propose a new class of eigenvector-like centrality measures that are computationally tractable and scalable in large-scale problems. First, let's provide some basics and notational conventions.

**Definition 1** *Cumulative Degree* of node $m$, $CD_m$ is defined as degree sum of the nodes in the neighbor set of $m$

$$CD_m = \sum_{i \in N_m} d(i)$$

Remark: In contrast to the so-called degree measure, this measure provides information about relational power of the node within its neighbors. In wireless network setting, this measure is easily computed with information available to each node within its *Personal Area Network* (PAN).

**Definition 2** Higher order Cumulative Degree of node $m$, $CDn_m$ is defined as degree sum of the nodes in the neighbor set of m up-to layer n along spanning tree $S_m$

$$CDn_m = \sum_{i_1 \in N_m} \cdots \sum_{i_n \in N_{i_{n-1}}} d(i)$$

**Definition 3** *Distributed Cumulative Degree* of node $m$, $DCD_m$ is defined as degree sum of the nodes in the neighbor set of $m$ along spanning tree $S_m$

$$DCD_m = \sum_{S_m} \cdots \sum d(i)$$

**Definition 4** *Discounted Distributed Cumulative Degree* of node $m$, $D^2CD_m$ is defined as discounted sum of the node degree of the nodes in neighbor set of $m$ along spanning tree $S_m$

$$DCD_m = \sum_{S_m}[\alpha_0 d(i) + \ldots \sum \alpha_n d(j)]$$

Where $n \in \mathfrak{R}$ is length of the spanning tree.

### 3.2. Performance Analysis

This section investigates the performance of the proposed measure. We have considered two datasets *Iranian Railways* (Fig. 2) and *Bucky ball* (Fig. 3) and then compare the result with other standard measures; *Closeness*, *Betweeness* and *Eigenvector* measure. These measures have several desired attributes, and our aim is to determine how well the proposed measures behaves under different network flow or how it scores nodes based on degree distribution using Railways sparse graph and the Bucky ball graph with strong connections.

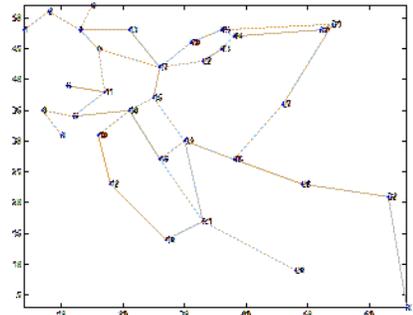

**Fig. 2 Iranian Railways**

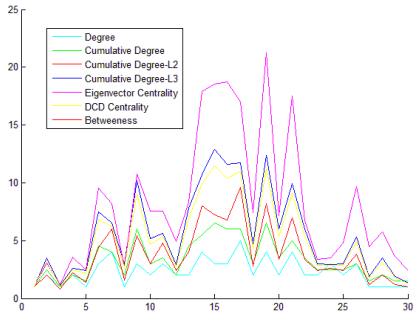

**Fig. 3 Comparison between standard *Betweeness* and *Eigenvector* centrality measure with the *Cumulative Degree* measure**

Fig. 3 shows that the higher order cumulative degree measures are able to represent desired properties like that of the eigenvector centrality in a distributed form while node degree measure fails in some nodes to provide effective information.

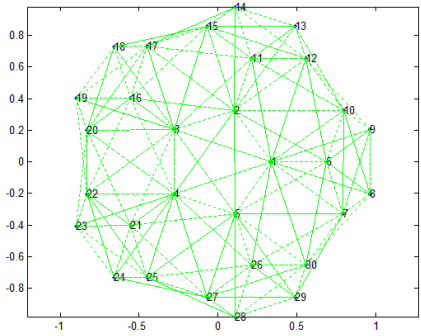

**Fig. 3 Bucky ball**

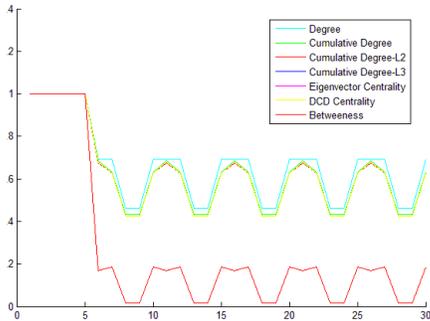

**Fig. 4 Comparison between standard *Betweeness* and *Eigenvector* centrality measure with the *Cumulative Degree* measure**

In Fig. 4, comparison represents a meaningful relation between cumulative degree measure and eigenvector centrality. As shown in Fig. 5, by expanding computation of the measure to higher order layer along spanning tree, the approximation error decreases below one percent.

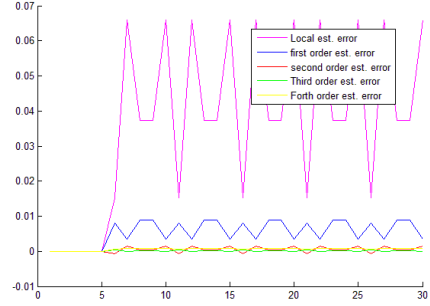

**Fig. 5 comparison error between *Cumulative Degree* measure and *Eigenvector* centrality measure**

### 3.3. Directed Consensus Algorithm

In this section, we introduce Directed Consensus algorithms that provide fastest mixing time by means of its directed and weighted information sharing mechanism. In a weighted directed consensus algorithm

$$x_i(t+1) = w_{ii}x_i(t) + \sum_{j \in N_i} w_{ij}x_j(t)$$

Weight design is performed by the following rule

$$w_{ij}(t) = \begin{cases} \dfrac{1}{\max(D_i, D_j)*(1+D_i-D_j)} & (i,j) \in E \\ 1 - \sum_{k \in N_i} w_{ik} & i = j \\ 0 & otherwise \end{cases}$$

Where D is one of the degree measures and communication between neighbors occurs in time-varying timeslots where its period is proportional to $1/w_{ij}$.

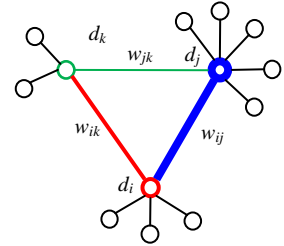

**Fig. 6 Weight design based on the cumulative degree measure**

Remark: Directed Consensus algorithm is inspired from biological systems and in particular from leaf venation patterns. The vascular pattern in leaves brings optimality and robustness to the corresponding directed consensus algorithm, even in time-varying networks.

Remark: It can be shown that the related weighted matrices with the proposed communication strategy satisfy the Para-contracting conditions and then the proposed algorithm converges [14].

# 4. APPLICATIONS

Management and optimization of Large-scale systems such as power systems, oil and gas transmission networks require efficient distributed algorithms to share information within network in an optimal and robust way. In a multi-agent setting, consensus algorithms used to share information between agents to reach a consensus and then plan independently. This information sharing scheme provides an average aggregation between multiple agents to coordinate their behavior with respect to neighbors.

Performance optimization and crisis management in Gas transmission networks has been recognized as a challenging problem and this problem have investigated from different disciplines. In this section, we present the application of the proposed measure augmented with a consensus scheme to share information and then coordinate compressor station activities and operation of remote control valves. This problem involves minimization of supply and demand gap as well as fuel consumption, and determining fault tolerance set points for safe operation of compressor station in the face of a disaster.

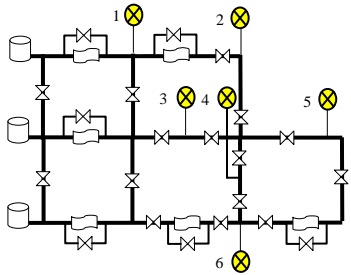

**Fig. 7 A part of Gas Transmission Network**

In this section, our approach is based on the consensus-seeking technique except to the aggregation quantities and the way information directed between agents. Fig. 7 shows a part of the *Iranian Gas Transmission Network* with three refineries that produces natural gas, and then transmitted by pipe-lines. Six compressor stations are used to maintain required pressures and flows for seven major consumers (e.g. cities and so on). Control valves are controlled by nearest station and are used to direct the network flow. *Simulink-Stateflow®* is used for hybrid system modeling in which each pipe segment has three operating modes; operating, break and leak. For operating mode, transient flow in pipe ducts is represented by transfer function models (see Fig. 8). In the same way, compressor stations may take advantage of three different modes; operation, recycling and shut-down (see Fig. 9). Control valves are also modeled by the following modes; open, close and fail (no control).

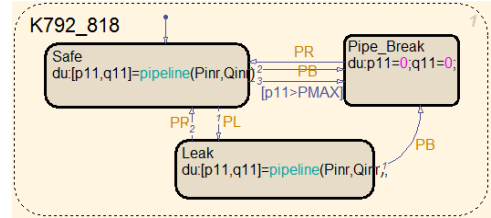

**Fig. 8 Hybrid transient model of a *Pipe Duct***

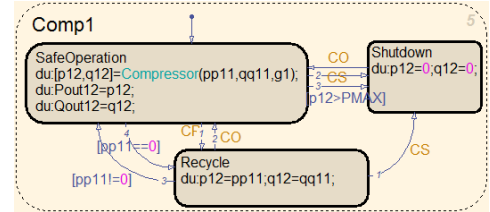

**Fig. 9 Hybrid model of *Compressor Station***

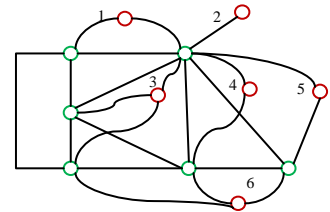

**Fig. 10 Network Graph of Fig. 7**

The main objectives of the control system are safety (fault-tolerant), optimization and crisis management. In the optimization case, a trade-off problem between different agents is solved based on the biased consensus algorithm. To maintain a safe and fault-tolerant operation and to recover network from major failures (like when a pipe-segment or compressor station break), directed consensus algorithm is used to provide new operating points for stations just in a few iterations.

In Table.1, some results of optimization algorithm are presented. In this case, we have used the directed consensus algorithm while aggregation between compressor stations and consumers achieved based on the weighted errors between (randomly generated) initial values and requested pressures.

| Units/Parms | Desired Pressure (psi) | Achieved Pressure (psi) | Decision Power in Optimization (D) |
|---|---|---|---|
| Consumer1 | 650 | 672 | 0.072 |
| Consumer2 | 812 | 803 | 0.053 |
| Consumer3 | 750 | 788 | 0.127 |
| Consumer4 | 640 | 664 | 0.089 |
| Consumer5 | 695 | 686 | 0.068 |
| Consumer6 | 730 | 719 | 0.116 |

**Table 1 Distributed optimization of gas transmission network**

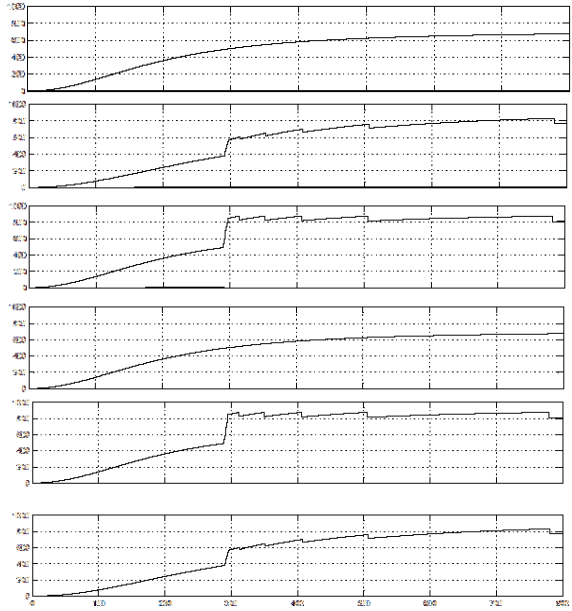

**Fig. 11 Different pressures of consumers (psi)**

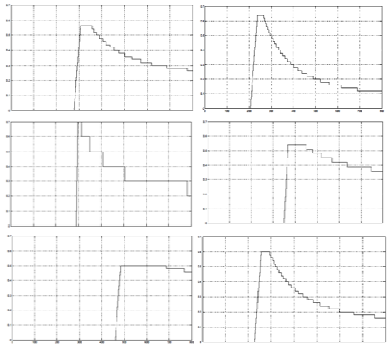

**Fig. 12 Compressor Station Ratios (0 to 0.7)**

## 5. CONCLUSION

In this paper, a new class of distributed centrality measures is introduced. These measures can be served as routing or scheduling tools in distributed algorithms. In recent years, consensus algorithms have shown their applicability and efficiency in distributed decision and control problems. We have shown that the proposed measures can be used to provide faster and more robust consensus between agents. In particular, we also investigate a routing scheme that is used to direct information within network. However, the convergence and performance analysis of the proposed measure require further considerations where addressed in our extended report [14].